# The Serendipitous Discovery of a Group or Cluster of Young Galaxies at $z \simeq 2.40$ in Deep HST[1] WFPC2 Images

S. M. Pascarelle[2,3] , R. A. Windhorst[2,4], S. P. Driver[2], E. J. Ostrander[3]

Dept. of Physics & Astronomy, Arizona State University, Tempe, AZ 85287-1504

Email:smp@deltat.la.asu.edu;raw,spd,ejo@cosmos.la.asu.edu

and

W. C. Keel

Dept. of Physics & Astronomy, University of Alabama, Tuscaloosa, AL 35487-0324

Email:keel@bildad.astr.ua.edu

---

## ABSTRACT

We report the serendipitous discovery of a group or cluster of young galaxies at $z \simeq 2.40$ in a 24-orbit HST/WFPC2 exposure of the field around the weak radio galaxy 53W002. Potential cluster members were identified on ground-based narrow-band redshifted Ly$\alpha$ images and confirmed via spectroscopy.

In addition to the known weak radio galaxy 53W002 at $z=2.390$, two other objects were found to have excess narrow-band Ly$\alpha$ emission at $z \simeq 2.40$. Both have been spectroscopically confirmed, and one clearly contains a weak AGN. They are located within one arcminute of 53W002, or $\sim 0.23 h_{100}^{-1}$Mpc ($q_o=0.5$) at $z \simeq 2.40$, which is the physical scale of a group or small cluster of galaxies. Profile fitting of the WFPC2 images shows that the objects are very compact, with scale lengths $\simeq 0\farcs 1$ ($\simeq 0.39 h_{100}^{-1}$kpc), and are rather faint (luminosities <L$^*$), implying that they may be sub-galactic sized objects. We discuss these results in the context of galaxy and cluster evolution and the role that weak AGN may play in the formation of young galaxies.

*Subject headings:* galaxies: clusters: general—galaxies: distances and redshifts—galaxies: evolution—galaxies: formation—quasars: general

## 1. Introduction

To gain a better understanding of galaxy formation we must observe a representative sample of galaxies over a wide range of lookback time. The difficulty, however, lies in the detection of a complete sample of galaxies with spectroscopic redshifts in excess of 1.0. Searches based on absorption lines seen in the spectra of background quasars, *e.g.*, Mg II or damped Ly$\alpha$ (Elston *et al.* 1991; Pei, Fall, & Bechtold 1991; Turnshek *et al.* 1991; Lowenthal *et al.* 1991 & 1995, hereafter L95; Wolfe *et al.* 1992; Macchetto *et al.* 1993, hereafter M93; Francis *et al.* 1995; Steidel, Dickinson, & Persson 1995), as well as serendipitous discoveries (Djorgovsky *et al.* 1985; Steidel, Sargent, & Dickinson *et al.* 1991), have so far yielded relatively few high-$z$ galaxies. Since most powerful *nearby* radio galaxies are luminous ellipticals, which often occur in groups or clusters, it may be possible to assemble a sample of galaxies through their association with known high-$z$ radio galaxies or QSOs (McCarthy *et al.* 1987; Lilly 1988; Hu *et al.* 1991; Windhorst *et al.* 1991, hereafter W91; Dressler *et al.* 1993). The opportunity to examine *several* galaxies at the same redshift in such a high-$z$ group would be extremely valuable to the study of galaxy evolution.

In this *Letter* we report the serendipitous discovery and consequent follow-up of a group or cluster around 53W002, a known weak steep-spectrum compact radio source at $z$=2.390 (W91). In §2 we present the observations, including both narrow-band imaging and spectroscopy, which confirm our initial discovery. We give our results in §3 and discuss their implications in §4.

## 2. Observations

### 2.1. The HST/WFPC2 Images

In May–June 1994, we obtained a 24-orbit exposure of 53W002 and its surroundings with the refurbished HST/WFPC2. The 0''.07 resolution images of 53W002 confirmed that ~20±4% of its central flux comes from an unresolved point source (Windhorst & Keel 1995, hereafter WK95). Its AGN is surrounded by a clearly extended and fairly symmetric $r^{1/4}$-like light profile with $r_e \simeq 1''.1 \pm 0.1$ ($\simeq 4.3 h_{100}^{-1}$ kpc), consistent with that of nearby *luminous* early-type galaxies. (For details, see W91; Windhorst, Mathis, & Keel 1992, hereafter WMK92; Windhorst *et al.* 1994; WK95.)

Figure 1 (Plate 1) shows a color image constructed from the 5.7 hr $V_{F606W}$ and $I_{F814W}$ WFPC2 images for the central portion of this field. Potential cluster members, chosen on the basis of their similarities in appearance to the serendipitously discoverd object "A," are labelled "B" through "J." Object "4" is from W91 and WMK92, and is discussed later. Details of the WFPC2 observations are given by WK95 and Driver *et al.* (1995). The dark HST sky at the North Ecliptic Pole allowed galaxies to be imaged down to V$\simeq$27 and I$\simeq$26 mag, while the few stars in the images can be detected down to V$\simeq$28.5 mag (see WK95).

### 2.2. MMT Spectra

As part of routine spectroscopic followup on the brighter galaxies in the field, we serendipitously discovered a blue compact object ("A" in Fig. 1) with $V_{F606W}$=23.07 mag at the same redshift as 53W002 to within ~400km/s. Its spectrum shows Ly$\alpha$, N V, C IV, and possibly O IV, He II, and C III emission lines at $z$=2.397, some of which have broad wings indicative of a weak AGN component (see Fig. 2$a$). The spectrum was obtained with the Blue Spectrograph at the MMT using a 300 gpm grating giving ~9Å resolution, and has been smoothed to ∼18Å resolution to increase signal-to-noise. In April 1995, Ly$\alpha$ and possibly C IV at $z$=2.393 were found in a short 1800 sec exposure of object "B" (Fig. 2$b$), a cluster candidate discovered in our narrow-band Steward images. The spectrum was obtained with the MMT Red Spectrograph and low resolution (∼20Å) 150 gpm grating.

### 2.3. Steward 90-inch $B$-band and Narrow-band Redshifted Ly$\alpha$ Images

In order to search for other potential cluster members, a narrow-band filter centered at 4130Å (Ly$\alpha$ at $z \simeq 2.40$) was quickly constructed by D. Marcus for use at the Steward Observatory 90-inch telescope. The blue-sensitized Loral 800×1200 CCD was used to image a 3′×4′.5 field at 0″.225 per pixel in both the "Ly$\alpha$" and Johnson $B$ filters. Due to poor weather, we were only able to obtain two 3600 sec exposures in Ly$\alpha$ and one 720 sec exposure in $B$. These two images are displayed in Figure 3 (Plate 2) and are equally deep despite their different exposure times, given the relative widths and throughputs of the two filters (∼150Å FWHM and ∼60% transmission for Ly$\alpha$ versus ∼1100Å FWHM and $\gtrsim$90% transmission for $B$). A deeper (6600 sec) Steward 90-inch $B$-band exposure of a 3′ × 3′ area of the same field was taken in 1990 by W91. The three spectroscopically confirmed $z \simeq$2.40 objects show obvious excess flux at 4130Å and are indicated in Figure 3.

In addition to the HST and Steward images, we also have accurate Gunn $gri$ photometry for all candidates from deep Palomar 200-inch $Fourshooter$ images (including objects "3" and "4" from W91, which were not detected significantly in the recent 90-inch Ly$\alpha$ images). We predicted $B$ magnitudes from the $gri$ fluxes following W91, and used them to calibrate the 1990 and 1994 Steward 90-inch $B$-band images. We then calculated weighted averages of the 1990 and 1994 $B$ magnitudes which are more accurate than those of the shallow 1994 exposure alone. All photometry for the cluster members is given in Table 1.

## 3. Results

We have updated the (4130Å–$B$) versus ($B$–$g$) color-color diagram of W91 with photometry for *all* objects surrounding 53W002 (Fig. 4). Because the 4130Å band is completely contained within the FWHM of the Johnson $B$-band, this diagram can locate objects with significant emission or absorption at 4130Å. A featureless power law ($F_\nu \propto \nu^{-\alpha}$) is indicated by the dotted line with tick-marks indicating integer values of $\alpha$. The three spectroscopically confirmed $z \simeq 2.40$ objects lie significantly below the line, and objects "3" and "4" are borderline candidates (based on previous photometry, W91). Multi-epoch photometry suggests that object "A" is variable on the timescale of years which explains its unreasonable spectral index, placing it in the lower right portion of Figure 4. It decreased in brightness by almost one magnitude in the Steward 90-inch $B$-band images from 1990 to 1994, which is well beyond our largest possible photometric errors. Such variability is consistent with its Seyfert-like spectrum (Fig. 2).

Light profiles were generated, as for 53W002 in WMK92 & WK95, for objects "A" and "B," and are given in Figure 5. Object "A" and the bright knot of object "B" are so compact that it was necessary to subpixellate their WFPC2 images five times (using biquintic interpolation). Effective radii were measured by assuming $r^{1/4}$ profiles, although exponential disks give similar results since their scale lengths are only ∼1-2 pixels. Model profiles were then generated for a range of effective radii (following Keel & Windhorst 1993), convolved with a similarly subpixellated star, resampled to real WFC pixels, and then subpixellated again to simulate the process used on the real data. Object "A" appears exactly stellar at small radii, but is somewhat larger than the star PSF beyond that. Its profile dips below the PSF at $r \gtrsim 7$ pix (0″.7) due to its location in the 'pyramid-shadow' region on the WFPC2 CCDs, while the reference star was much further from the chip edge. Its 'best-fit' effective radius is $r_e \lesssim 0″.1 \pm 0″.02$. Object "B" appears to be in an interacting (or otherwise disrupted) system, or possibly has some reflected AGN light on one side from the compact component. Model fitting of the compact knot gives $r_e \simeq 0\rlap.{''}1 \pm 0\rlap.{''}02$. Note that the profile appears extended because it runs into its neighboring companion or cloud at $r \gtrsim 4$ pix (see Fig. 1).

## 4. Discussion

These findings raise several questions, the most interesting of which is why the confirmed cluster members (other than the radio source), after subtracting their likely point-source contributions, are so small and faint. Table 1 contains the half-light radii, point-source contributions, and absolute $V$ magnitudes after point-source subtraction for the three $z \simeq 2.40$ cluster members. W91 found that at $M_v \simeq -23.6$ (assuming a 35% point-source contribution), 53W002 is only as luminous as an (evolving) $L^*$ galaxy, whereas the more powerful radio galaxies at similar redshifts have luminous masses $\simeq$5-8 times greater (depending on the choice of cosmological parameters). The underlying stellar population in object "A," which is about two magnitudes fainter than 53W002, is thus most likely sub-$L^*$. This implies that perhaps we are seeing a young galaxy in the early stages of its developement, whose luminosity has not yet reached that of a typical $L^*$ galaxy at $z \simeq 2.40$. Object "B," which is as compact as object "A," but doesn't necessarily contain an AGN based on the current spectral data, is also $\lesssim L^*$ at $z \simeq 2.40$. Estimating the absolute $I$ magnitudes of these galaxies from Figure 6 of Casertano *et al.* (1995) also gives values below that of $L^*$ galaxies ($M_I \simeq -22.7$).

Several selection biases must be noted, the most important of which is the preferential detection of cluster members with significant Ly$\alpha$ emission. This implies that the cluster members were found only because they contain weak AGN. However, it has recently become clear that, despite issues of resonant scattering and destruction of Ly$\alpha$ photons by dust (Meier & Terlevich 1981; Hartmann *et al.* 1988; Terlevich *et al.* 1993), actively star-forming galaxies can in fact also have substantial Ly$\alpha$ emission (Calzetti & Kinney 1992; Valls-Gabaud 1993; L95). Several groups have now reported detections of Ly$\alpha$ emission from high-redshift galaxies, in which either the spatial extent or continuum shape suggest that it is produced by stellar photoionization rather than an active nucleus (Cowie & Lilly 1989; M93; Moller & Warren 1993; Giavalisco, Steidel, & Szalay 1994). Therefore, detectable Ly$\alpha$ emission does not always require an AGN.

Surface brightness selection effects caused by the small pixels in WFPC2, the cosmological $(1+z)^{3+\alpha}$ dimming, and the fact that we are observing at emitted wavelengths of 1800-2400Å mean that the compact appearance of the cluster candidates must be viewed with some caution. The galaxies surrounding compact regions of strong UV light could be significantly larger than our measured sizes if their surface brightnesses are low enough to be completely lost in the WFPC2 sky + read-noise, although we do attempt to avoid this problem by measuring half-light rather than isophotal radii.

Despite all these selection effects, it is tempting to speculate, in light of the present data, on the possibility that all galaxies start out with (weak) active nucleii. Will the compact, sub-L$^*$ objects of this cluster in fact someday become the L$^*$ galaxies we see at the present epoch? With respect to large scale structure formation, can such a young cluster be explained by Cold Dark Matter? We plan to address these questions in future work (Pascarelle, *et al.* 1995). Current plans to extend these observations include 39 orbits in Cycle 5 with HST/WFPC2 using filters F410M (redshifted Ly$\alpha$) and F450W (Johnson B) from which we will be able to locate compact Ly$\alpha$-emitting components at scales 5–10 times smaller than those in this paper.

In conclusion, we have shown substantial evidence for the existence of a group or cluster of young galaxies at $z \simeq 2.40$. The two confirmed members (other than the radio

– 9 –

galaxy) appear to be compact and sub-L$^*$, with smooth well-formed inner light profiles. At least two, and possibly all three, show evidence for a weak AGN, suggesting that the presence of an AGN *may* be an integral part of galaxy formation, at least for galaxies of a certain luminosity. We look forward to being able to address many of the concerns about selection biases discussed here and shedding some light on the existence of such compact and faint objects with the superior quality of the HST images.

We would like to thank the staffs of the MMT and Steward 90-inch for their assistance, David Marcus for designing the narrow-band Ly$\alpha$ filter on such short notice, and Steven Mutz for helping with the 90-inch observations. We also thank the referee, James Lowenthal, for his very useful suggestions and comments on the manuscript. We acknowledge support from HST grants GO.5308.01.93A(RAW) and GO.5308.02.93A(WCK).

Table 1: Photometry of Cluster Candidates

| Name | RA<br>Dec | $U$<br>(mag) | 4130Å<br>(mag) | $B$<br>(mag) | $V$<br>(mag) | $I$<br>(mag) | Gunn $g$<br>(mag) | Gunn $r$<br>(mag) | Gunn $i$<br>(mag) | $r_{hl}$<br>(") | % unresolved<br>point source | $M_v$<br>(mag) |
|---|---|---|---|---|---|---|---|---|---|---|---|---|
| 53W002 | 17:14:14.74<br>50:15:28.94 | 23.24<br>(0.36)[a] | 22.41<br>(0.19) | 23.37<br>(0.17) | 23.01<br>(0.08) | 22.48<br>(0.14) | 22.87<br>(0.07) | 22.83<br>(0.07) | 22.78<br>(0.10) | 1.1 | ∼20[b] | −23.9<br>(∼L*) |
| Object A | 17:14:11.27<br>50:16:08.74 | 22.57<br>(0.42) | 21.97<br>(0.17) | 23.15<br>(0.36) | 23.07<br>(0.14) | 22.30<br>(0.12) | 21.69<br>(0.06) | 22.29<br>(0.07) | 22.77<br>(0.16) | $\lesssim 0.1$ | ∼90 | −21.4[c] |
| Object B | 17:14:11.89<br>50:16:00.51 | 22.51<br>(0.39) | 21.36<br>(0.17) | 23.28<br>(0.22) | 23.17<br>(0.14) | 22.78<br>(0.18) | 22.79<br>(0.10) | 23.01<br>(0.13) | 23.00<br>(0.23) | 0.1 | ∼52 | −23.0[c] |

[a]Errors are the quadratic sum of the formal photometric errors (from photon statistics and the rms uncertainty in the surrounding sky fits) and the zeropoint error. The latter is approximately 0.2 mag in $U$, 0.15 mag in Ly$\alpha$, 0.12 mag in $B$, 0.07 mag in $V_{F606W}$ and $I_{F814W}$, and 0.05 mag in Gunn $gri$.

[b]from WK95

[c]Determined by using the distance modulus and $K$-correction of 53W002 (c.f., Table 5 in W91, based on Bruzual's 1983 spectral evolution models for a young stellar population at $z \simeq 2.40$).

## 5. Figure Captions

**Fig. 1.** (Color Plate 1) HST Cycle 4 color image of the field surrounding the weak radio galaxy 53W002 at $z$=2.390. The image was constructed from a 5.7 hr $V_{F606W}$ and a 5.7 hr $I_{F814W}$ WFPC2 exposure and is about $72''\times72''$ with $0\rlap.{''}0966$ pixels. Objects "A" and "B" are the spectroscopically confirmed cluster members at $z\simeq2.40$.

**Fig. 2.** (*a*) 2×2700 sec longslit spectrum of object "A" showing the emission lines of Ly$\alpha$, N V, C IV, and possibly O IV, He II, and C III at $z$=2.398. The shape of the continuum and the relatively narrow lines with broad wings are indicative of a weak AGN. (*b*) 1800 sec longslit spectrum of object "B" showing ly$\alpha$ and C IV, with possibly N V and He II at $z$=2.393.

**Fig. 3.** (Plate 2) (*a*) 2×3600 sec Ly$\alpha$ image of the field near 53W002 taken at the Steward 90-inch through a narrow-band redshifted Ly$\alpha$ filter (150Å FWHM) centered at 4130Å. (*b*) 720 sec broad-band Johnson $B$ image of the same field. The circled objects are cluster candidates based only on their apparent excess Ly$\alpha$ emission in the upper panel.

**Fig. 4.** (4130Å–$B$) versus ($B$–$g$) color-color diagram for all objects surrounding 53W002, based on both recent 90-inch data and older 90-inch and Palomar data (W91). The locus of featureless power-law spectra ($F_\nu \propto \nu^{-\alpha}$) is indicated by the dotted line with the crosses marking integer values of $\alpha$. Objects "A" and "B" are spectroscopically confirmed cluster members. The unusual location of object "A" in the diagram is most likely due to its variability.

**Fig. 5.** Light profiles for confirmed cluster members "A" and "B." Both are very compact and were analyzed by subpixellating their HST images five times (using biquintic interpolation).